\documentclass[aps,preprint]{revtex4}
\usepackage{graphicx}%
\usepackage{dcolumn}
\usepackage{amsmath}

\newcommand{\lessim} {\mathop{\,<\kern - 1.05 em \lower 1.ex \hbox {$\sim$}\,}}
\newcommand{\grtsim} {\mathop {\,> \kern - 1.05 em \lower 1.ex \hbox 
{$\sim$}\,}}

\begin{document}
\normalsize  \title{New SDW phases in quasi-one-dimensional systems   
dimerized in the transverse direction}

\author{Dra\v{z}en Zanchi}
\affiliation{Laboratoire de Physique Th\'eorique et Hautes Energies. 
Universit\'es Paris VI Pierre et Marie Curie -- Paris VII Denis Diderot, 
2 Place Jussieu, 75252 Paris C\'edex 05, France.}
\author{Aleksa Bjeli\v{s}}
\affiliation{Department of Physics, Faculty of Science, University of
Zagreb, POB 162, 10001
Zagreb, Croatia}

\widetext
\begin{abstract}

The spin density wave instabilities in the 
quasi-one-dimensional metal (TMTSF)$_2$ClO$_4$ are studied in the 
framework a matrix random phase approximation for intra-band and inter-band 
order parameters. Depending on the anion ordering potential $V$ which measures  
the lattice doubling in the transverse direction, two different 
instabilities are possible. The SDW$_0$ state at low values of $V$ 
is antiferromagnetic in $b$ direction and has the critical temperature
that decreases rapidly with $V$. The degenerated states SDW$_{\pm}$, 
stable at higher values of $V$, are superpositions of two magnetic orders, 
each one on its subfamily of chains. As $V$ increases the ratio between two 
components of SDW$_{\pm}$ tends to zero and the critical
temperature increases asymptotically towards that of SDW instability for a
system having perfect nesting and no anion order. 
At intermediate $V$, i.e. between the phases SDW$_0$ and SDW$_{\pm}$, 
the metallic state can persist down to $T=0$.


\end{abstract}

\bigskip
LPTHE/01-18 \hspace{106mm}May 2001
\bigskip

\maketitle

\bigskip



\newpage

\begin{figure}[htbp]
\begin{center}
\setlength{\unitlength}{1cm}
\begin{picture}(7,3)
\put(-3.5,-7.5){\includegraphics[width=14cm]
{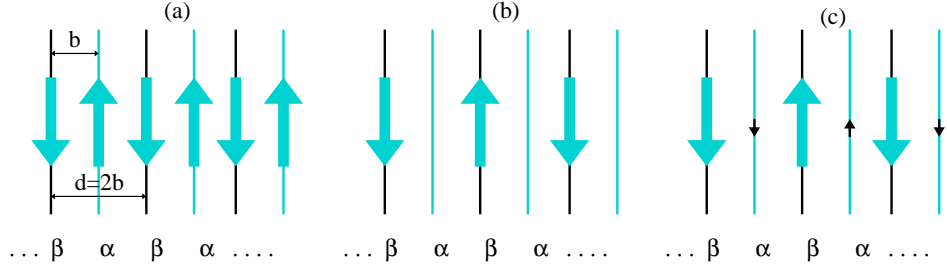}}
\end{picture}
\caption{Magnetic pattern in direction perpendicular to chains for three
characteristic cases of transverse SDW modulation: (a) the limit of weak
anion potential $V$ with $p=\pi/b$; (b) the
limit of large  $V$ with $p=\pi/d$; (c) $V/t_b$  moderate with $p=\pi/d$
transverse  SDW order.}
\label{abab}
\end{center}
\end{figure}

Among the quasi--one--dimensional systems (TMTSF)$_2X$ with the generic 
name Bechgaard salts the material X=ClO$_4$ has a particular status. 
If relaxed, it is a superconductor at ambient pressure and 
develops a field--induced spin--density--wave (FISDW) cascade for the magnetic 
fields $B$  above  3 Tesla, well above the critical field for 
superconductivity. At even higher
values of $B$  it possesses a still controversial type of SDW 
ordering.\cite{Chaikin}
If the sample is cooled more rapidly, the SDW order can persist even down to 
zero magnetic field. The critical temperature of this zero--field 
transition increases with the cooling rate  up to the value of 
about 5.5 K for  quenched  samples.\cite{Qualls00,Matsunaga} 
It is believed that this strong dependence of the SDW ordering on the cooling 
rate
and its absence in carefully relaxed samples are 
related to the fact that (TMTSF)$_2$ClO$_4$ in its ground state  
acquires a long range order of anion molecules  ClO$_4$.
This ordering sets up at the temperatures up to about 24K, 
inducing a doubling of the unit cell perpendicularly to chains,
with the new lattice constant $d=2b$, $b$ being the inter--chain distance, as shown on 
Fig. \ref{abab}. By varying the cooling rate one realizes 
a tunable unit cell doubling which in turn strongly influences the SDW 
ordering.

The usual and the most direct
measure of this doubling on the electron band  properties is the band 
splitting $V$ of $\alpha$ and $\beta$  sub-families of chains 
(Fig.  \ref{abab}).\cite{Lepeleven} 
We take that $V=-|V|$ on $\alpha$ chains and $V=|V|$ on $\beta$ chains.
It was already pointed out that due to this splitting 
the two SDWs shown on  Figs. \ref{abab}(a) and \ref{abab}(b) are in 
competition.\cite{Chaikin,Kishigi_etc,Kishigi98}
Realizing that these SDWs cannot be treated independently (i.e. additively) 
as it was done in the
existing bibliography, we develop a generalized matrix method within the random
phase approximation (RPA). In this Letter we concentrate on 
(TMTSF)$_2$ClO$_4$ in the absence of magnetic field, and determine the phase 
diagram in  whole range of values of the band splitting $V$, particularly 
for the physically relevant intermediate values, not properly covered in the 
previous analyses.

As it is well known  the Coulomb interaction between electrons leads to a SDW 
ordering via the mechanism of Fermi surface nesting.\cite{Nesting}
If the chains were equivalent the nesting would lead to a SDW order with the 
magnetization modulation characterized by the wave-vector $(2k_F,\pi/b)$ as 
shown on Fig. \ref{abab}(a). 
 One can imagine that in the 
limit of strong dimerization the nesting mechanism is again in action, but now 
separately for bands of $\alpha$ and $\beta$ sub-families of chains from 
Fig. \ref{abab}, and conclude that the critical temperature  for each
subfamily is  of the same 
order as that for equivalent chains  provided the new nesting is again (close 
to) perfect. The new SDW then has doubled period $2d=4b$ in the $b$ direction, 
as shown on Fig.\ref{abab}(b)  for the subfamily $\beta$. 
Our analysis confirms that this indeed happens 
in the limit of large anion potential,  $V>>t_b$, where $t_b$ is the 
interchain hopping integral. However for
intermediate values of $V$ the situation is more subtle due to the mixing of 
two SDWs.  There, a new type of SDW dominates the response. 
It is a combination of two SDW orders on two different sublattices,
each one forming an  antiferromagnetic pattern in $b$ direction,
as shown  schematically on Fig. \ref{abab}(c). We show that then the 
critical temperature is reduced or even completely suppressed.

All modulations from Fig. \ref{abab} can be followed in terms of 
homogeneous and alternating SDW amplitudes defined on the dimerized
lattice
\begin{equation}\nonumber
{\bf M}_{h}(x,n)={\bf M}_{\alpha}(x,nd) + {\bf M}_{\beta}(x,nd +d/2)=
\frac{1}{2}[{\bf M}_{11}(x,n) + {\bf M}_{22}(x,n)]
\end{equation}
\begin{equation}
\label{amp}
{\bf M}_{a}(x,n)={\bf M}_{\alpha}(x,nd) - {\bf M}_{\beta}(x,nd +d/2)=
\frac{1}{2}[{\bf M}_{12}(x,n) + {\bf M}_{21}(x,n)],
\end{equation}
represented for later convenience in terms of  SDW amplitudes 
on $\alpha(\beta) $ chains
${\bf M}_{\alpha(\beta)}= c_{+\alpha(\beta)}^{\dagger}
{\bf \sigma}c_{-\alpha(\beta)}$, and in terms of 
${\bf M}_{ij}= \Psi_{+,i}^{\dagger}{\bf \sigma}\Psi_{-,j}$ 
(with $i,j = 1,2$) which are the SDW amplitudes in terms of the bond/antibond states
 \begin{equation} \label{ba}
\Psi_{f, 1(2)}(x, n) = \frac{1}{\sqrt{2}}[c_{f,\alpha}(x, nd) 
\pm c_{f,\beta}(x, nd+d/2)],
\end{equation}
with $f=+,-$. Here $c_{\pm\alpha(\beta)}$ are two component field operators 
for the right (left) Fermi sheet, and ${\bf \sigma}$ is the Pauli operator 
in the spin space. 
Note that operators defined in the bond/antibond basis 
($\Psi$, ${\bf M}_{ij}$ and ${\bf M}_{h(a)}$) depend only on the cell index 
$n$, while the local operators $c$ and ${\bf M}_{\alpha(\beta)}$ depend also 
on the position within the unit cell. In particular this means that 
$c_{f,\alpha}(x,nd+d/2)=c_{f,\beta}(x,nd)=0$ and 
${\bf M}_{\alpha}(x,nd+d/2)={\bf M}_{\beta}(x,nd)=0$.

Below $T_c$ both ${\bf M}_h$ and ${\bf M}_a$ can attain non-zero
mean  values, with some specific common wave vector ${\bf q}=(k,p)$ 
and with the common orientation along some easy axis, say $\hat{z}$.
The staggered magnetization is then
\begin{equation} \label{mag}
m_z(x,R_{\perp})=
(\Delta _h \pm  \Delta _a)\cos{\left[ (2k_F+k)x+
pnd \right]} \; ,
\end{equation}
where the upper and lower sign stay for $\alpha \,(R_{\perp}=nd)$ and 
$\beta \,(R_{\perp}=nd+d/2)$ chains respectively, and $\Delta _{h/a}$  are 
the mean values of the critical 
Fourier components $(k,p)$ of ${\bf M}_{h/a}$. With  $\Delta _a\neq 0$, 
$\Delta _h= 0$ and  $p=0$ we get a simple antiferromagnetic order in 
$b$ direction [Fig. \ref{abab}(a)], like in the case of indistinguishable 
chains. We call this state SDW$_0$. Another limit with doubled periodicity 
[Fig. \ref{abab}(b)] is realized for $\Delta _h=\Delta _a$ or 
$\Delta _h=-\Delta _a$ with $p=\pi/d$. In this case the SDW with
antiferromagnetic modulation in the transverse direction can be realized 
either 
on the chains $\alpha$  or on the chains $\beta$, each ordering having its own
longitudinal modulation, with respective wave-numbers $2k_F+k_1$ and 
$2k_F-k_1$.

The intermediate   ordering from Fig.\ref{abab}(c) 
is realized for $|\Delta_h/\Delta _a| > 1$. The corresponding ordered 
states with larger ($2k_F+k_1$) and  smaller ($2k_F-k_1$) value of the 
longitudinal wave number will be denoted by SDW$_+$ and SDW$_-$ respectively.  
Fig. \ref{abab}(c) shows the ordering with $\Delta_h/\Delta _a<-1$, i. e. 
with the larger SDW amplitude on $\beta$ chains. Its longitudinal
modulation has the wavenumber $2k_F-k_1$.

Written in terms of the Fourier transforms of the operators 
(\ref{amp}) in the perpendicular ($b$) direction with the Brillouin zone 
defined by
($-\pi/d < q < \pi/d$), the interaction term of the Hamiltonian reads
 \begin{equation} \label{hint}
H_{I} = -\frac{U}{2N} \int dx \sum_{q}[{\bf M}_{h}^{\dagger}(x, q)
{\bf M}_{h}(x, q) +  {\bf M}_{a}^{\dagger}(x, q)  {\bf M}_{a}(x, q)],         
\end{equation}
while the band contribution, written in terms of bond/antibond operators 
defined by eq.(\ref{ba}), reads
\begin{equation} 
H_0= \frac{1}{2N}\sum_{\sigma}\sum_{p=-\pi/d}^{\pi/d} \int dx\;
\Psi ^{\dagger}(x,p) \biggl[ iv_F\rho_3  \partial _x +  
 2\tau_3t_b\cos{\frac{pd}{2}}
+2t_b'\cos{pd}
-V\tau _1\biggr] \Psi (x,p) 
\; .
\label{hkin}
\end{equation}
Here $\Psi ^{\dagger}(x,p)= (\Psi_{1} ^{\dagger}(x,p), 
\Psi_{2} ^{\dagger}(x,p))$, $\rho$'s and $\tau$'s are Pauli matrices in 
left-right and bond-antibond indices respectively, $t_{b}^{'}$ is the 
parameter which introduces an imperfect nesting in the absence of the anion 
ordering, and $2N$ is the total number of chains. 
Also, being interested only in the SDW channel, we keep here only the forward 
part of the presumably local (intrachain) electron-electron coupling in 
$H_{I}$, omitting also the weak Umklapp term due to the existing slight 
lattice dimerization along the chain.

Following the above qualitative arguments,
it appears natural to construct the RPA formalism in the ${\bf M}_{a,h}$  
basis. However, although  the interaction term $H_{I}$ is separated in 
${\bf M}_{a}$ and  ${\bf M}_{h}$, these SDW  amplitudes are coupled through 
the non-diagonal part of the kinetic term (\ref{hkin}), induced by anion 
ordering. Since the electron propagators  in the bond-antibond basis are not 
diagonal, the RPA equations for ${\bf M}_{a(h)}$ amplitudes are also coupled.  
On gets the static SDW susceptibility in the form of a 2x2 matrix, 
\begin{equation} \label{matrix}
[\chi({\bf q})]\equiv
\left(\langle
M_{i}({\bf q})M^{\dagger}_{j}({\bf q})\rangle\right)
=D^{-1}\left(
\begin{array}{cccc}
 (1-U\chi_{aa})\chi_{hh}+U(\chi_{ha})^2 &  \chi_{ha}\\ 
    \chi_{ha}   & (1-U\chi_{hh})\chi _{aa}+U(\chi_{ha})^2
\end{array}
\right) ,
\end{equation}
where indices $i$ and $j$ stand for $h$ and $a$, $M_i$ is the projection of 
${\bf M}_i$ onto the 
spin quantization axis, $D({\bf q})\equiv {(1-U\chi_{hh})(1-U\chi_{aa})-
U^2(\chi_{ha})^2}$, and 
$\chi_{hh}, \chi_{aa}$ and $\chi_{ah}=\chi_{ha}$  are corresponding bare 
(Hartree-Fock) 
susceptibilities.

The diagonalization of the matrix (\ref{matrix}) leads to 
\begin{equation} \label{diag}
[\chi({\bf q})]=\mbox{diag}\left(\frac{\tilde{\chi}_1}{1-U\tilde{\chi}_1},
\frac{\tilde{\chi}_2}{1-U\tilde{\chi}_2}\right)\; ,
\end{equation}
with 
\begin{equation} \label{tilde}
\tilde{\chi}_{1/2}=\frac{1}{2}\left[\chi_{aa}+\chi_{hh}\pm
\sqrt{\left(\chi_{aa}-\chi_{hh}\right)^2+4(\chi_{ha})^2}\right]\;. 
\end{equation}
The SDW instability takes place for 
\begin{equation} \label{stoner}
1-U\tilde{\chi}_1(T_c,{\bf q}_c)=0, 
\end{equation}
${\bf q}_c$ being the wave vector at which $\tilde{\chi}_1({\bf q})$
has the maximum. The corresponding ratio of two SDW order parameters 
from eq.(\ref{mag}) is
 \begin{equation}\label{ratio}
\frac{{\Delta}_h}{{\Delta}_a}=\pm\sqrt{\frac{1-\eta}{\eta}} \;,
\end{equation}
with
\begin{equation} \label{eta}
\eta = \frac{1}{2} \left[ 1 + \frac{\chi_{aa}-\chi_{hh}}
{\sqrt{\left(\chi_{aa}-\chi_{hh}\right)^2+4(\chi_{ha})^2}}\right].
\end{equation}
The sign ambiguity in eq.(\ref{ratio}) 
comes from the freedom of choice of the unit 
cell: $(\alpha,\beta)$ or $(\beta,\alpha)$.

It is convenient for further considerations to write the critical SDW
susceptibility $\tilde{\chi}_1$ in the form
$\tilde{\chi}_{1}(k,p)=\frac{1}{2\pi v_F} 
\left[ \ln{\frac{\epsilon}{T}} + 
\chi_{\mbox{\small res}}(k,p)\right]
$,
where $\chi_{\mbox{\small res}}$ is always negative and contains all 
corrections to the pure logarithm. It can be derived after passing to the
electron basis that diagonalizes the kinetic part (\ref{hkin}) of the 
Hamiltonian. The result is 
\begin{eqnarray}\nonumber
\lefteqn{ \chi_{\mbox{\small res}}(k,p) = 
\Psi(\frac{1}{2}) -
\frac{1}{4} \biggl<\sum_{X,Y}\left(1+
4XY\alpha\beta \alpha_{p}\beta_{p}\right)
\Psi_{\mbox{\small XY}}
 \biggr> } &  \\ 
& +  & \frac{1}{4}\biggl[ \biggl< (\alpha^2 - \beta^2)
(\alpha^2_{p} - \beta^2_{p})
\sum_{X,Y}XY\Psi_{XY}\biggr> ^2 
 + 16\biggl< \alpha\beta\sum_{X,Y}X\Psi_{\mbox{\small XY}}
\biggr>^{2}\biggr]^{1/2}.
\label{res}
\end{eqnarray}
Here $\langle...\rangle$ means the average along the transverse momentum 
$(-\pi/d<p'<\pi/d)$, $\Psi$ is di-gamma function, $X$ and $Y$ take the values
$\pm 1$, $\Psi_{\mbox{\small XY}}\equiv \Re\Psi(\frac{1}{2} + 
i Q_{\mbox{\small XY}})$ with
$
Q_{\mbox{\small XY}} \equiv \frac{1}{4\pi T}\biggl[  v_F k
+ 2 t_{b}'\left[\cos(p' + p) + \cos p'\right] - 
X \Delta(p') - Y\Delta(p' + p)\biggr],
$
and $\Delta(p) \equiv \sqrt{V^2 + \left[2 t_b\cos(\frac{pd}{2})\right]^2}$.
$\alpha$ and $\beta$ are the coefficients defining the unitary transformation 
that diagonalizes the band term (\ref{hkin}), 
\begin{equation} \label{albe}
\left\{ \begin{array}{l} \alpha(p')\\
                         \beta(p')\\
\end{array} \right\}
= \sqrt{\frac{1}{2}\left[ 1 \pm 
\frac{2t_b \arrowvert\cos(\frac{p'd}{2})\arrowvert}
{\Delta(p')}\right]},
\end{equation}
with the convention $\alpha_{p} \equiv \alpha(p'+p)$
used to shorten writing in eq. (\ref{res}). The corresponding
dispersions of two sub-bands are given by
\begin{equation} \label{subb}
E_{\pm f}(k,p) = f v_F k + 2 t_{b}^{'} \cos(pd) 
\pm \Delta(p)\, .
\end{equation}

The critical temperature is now the solution of the equation
$
\ln\frac{T_c}{T_c^{0}}=\chi_{\mbox{\small res}}
(T_c,V,t_b,t_b';{\bf q}_c)\; ,
$
where ${\bf q}_c$ is always chosen to maximize $\chi_{\mbox{\small res}}$, and 
$T_c^{0}\equiv T_c(V=0,t_b'=0)= \epsilon \exp(- \frac{\pi v_F}{U})$ 
is the $t_b$-independent critical temperature for the asymptotic case of 
perfect nesting $(t_b'=0)$ and vanishing anion order $(V=0)$; $\epsilon$ is
a constant of the order of the bandwidth.

\begin{figure}[htbp]
  \begin{center}
    \setlength{\unitlength}{1cm}
    \begin{picture}(7,6.5)
        \put(-6,-8.5){\includegraphics[width=18cm]{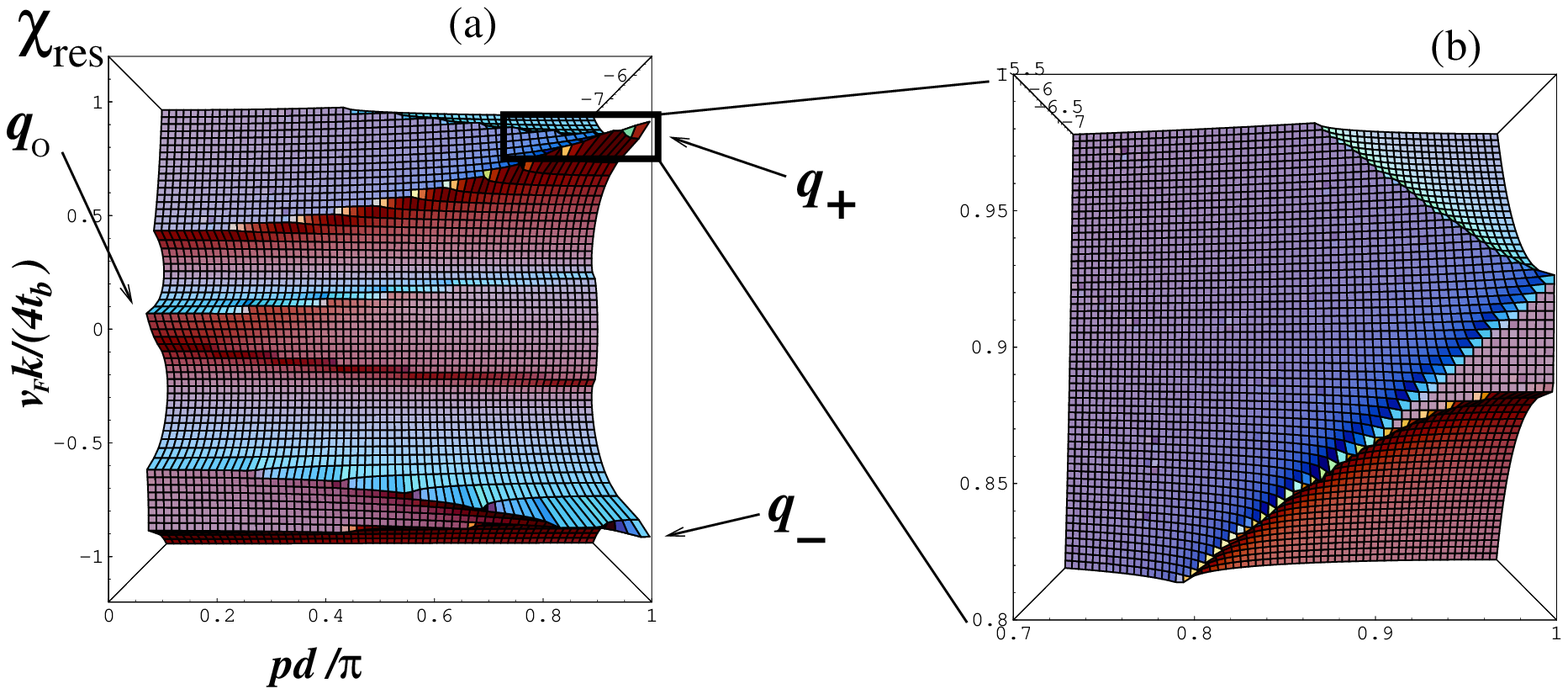}}
    \end{picture}
    \caption{The quantity $\chi_{\mbox{\small res}}$}
    \label{omega1}
  \end{center}
\end{figure}

The wave vector dependence of $\chi_{\mbox{\small res}}(k,p)$ is illustrated 
on Fig. \ref{omega1}a, in which we chose $t_b'/t_b = 0.1, V/t_b = 1.2$, and a 
very low temperature  $(T/t_b= 0.0004)$ in order to make the whole fine 
structure of this dependence visible. There is one peak at 
${\bf q}_0=(4t_b'/v_F,0)$, and two mirroring peak structures positioned 
approximately at ${\bf q}_{\pm}= ( k^{\pm}, \pi/d)$, with $k^{\pm}$ to be 
specified below. 

The peak at ${\bf q}_0$ corresponds to the phase SDW$_0$. It  reproduces the 
standard SDW instability appearing in the absence of the anion 
ordering.\cite{Nesting,Chaikin} Indeed it dominates over the peaks at 
${\bf q}_{\pm}$ for low values of $V/t_b$, as it will be
shown later in the discussion of the whole phase diagram. If in addition
the temperature is not too much lower than $t_b'$ this peak is smeared and  
is simply shifted to $k=0$, i.e. to the SDW$_0$ with a pure $2k_F$ longitudinal 
modulation.  One gets the two-band version of the well-known imperfect nesting 
induced commensurate--incommensurate transition \cite{Nesting}, with 
$\eta\approx 1$, i. e. with the SDW$_0$ order very close to an antiferromagnet 
in the transverse direction as shown on Fig. \ref{abab}(a). Here SDW$_0$ is
stabilized as an interband process: the wave vector 
${\bf Q}_{0}=2k_F\hat{a}+{\bf q}_0$ is the best nesting vector connecting 
the left Fermi sheet of the sub-band ``+'' with the right Fermi sheet of the 
sub-band ``-'', whose dispersions are given by eq. (\ref{subb}). Note that in 
this range of wave vectors the anion potential $V$ is also, beside the
original imperfect nesting $t_b'$, the source of deviation from the purely
logarithmic susceptibility. 
It becomes clear from the expression (\ref{res}) which after putting 
$k=0, \, p=0$ and $t_b'=0$ reads
\begin{equation}
 \chi_{\mbox{\small res}}(0,0) \approx 
\biggl< \frac{V^2}{V^2 + \left[2 t_b\cos(\frac{p'd}{2})\right]^2}
\biggl[\Psi\left(\frac{1}{2}\right) - \Re \Psi \left( \frac{1}{2} + 
\frac{i\sqrt{V^2 + \left[2 t_b\cos(\frac{p'd}{2})\right]^2}}{2\pi T}
\right)\biggr]\biggr>.
\label{resv}
\end{equation}
 Note that one gets the same expression after putting $k=0, \, p=0$ and 
$t_b'=0$ into the result of Miyazaki et al \cite{Miyazaki98}. 
For this particular case the off-diagonal element of the SDW susceptibility 
(\ref{matrix}) vanishes, so that the matrix approach reduces to the standard
scalar one.

The peak at ${\bf q}_{+}$ on Fig. \ref{omega1}a is in fact a plateau  
delimited on the axe $p=\pi/d$ by the wave numbers
$k^+_{1}=(\sqrt{V^2+4t_b^2}+V)/v_F$ and $k^+_{2}=2\sqrt{V^2+2t_b^2}/v_F$,
as shown by the enlarged picture of this part of Brillouin zone  on  
Fig.\ref{omega1}(b). This range of wave vectors is the intra-band 
nesting range 
for the sub-band $E_{-}$ in eq.(\ref{subb}). Equivalently, the range  
${\bf q}_{-}$ has the same role for the sub-band $E_{+}$. The complete
degeneracy of these two ranges would be raised after inclusion of higher 
corrections to the longitudinal linear dispersion in the one-electron 
spectrum assumed here. E. g. for the band--filling lower then one half 
the dominant SDW ordering is realized on the subfamily of chains which is 
higher in energy because it has a lower Fermi velocity and, consequently, a 
higher density of states.  In that case, the minus sign is to be
chosen in eq. (\ref{ratio}).

The wave numbers $k^+_{1}$ and $k^+_{2}$ delimit the region where after the 
translation by the wave vector ${\bf Q}_{+}=2k_F{\bf \hat{a}}+{\bf q}_+$ 
the left and right Fermi sheets cross each other, from the region where these 
sheets are not in contact. The difference $\Delta k=k^+_1-k^+_2$ measures the 
strength of effective imperfect nesting for the intraband SDW$_+$ ordering. 
In the limit of large anion splitting, $V\gg t_b$, it reduces to 
$v_F\Delta k\approx t_b^4/V^3 \equiv 8\epsilon _{\mbox{eff.}}$. 
In the language 
of the standard nesting model for SDW ordering \cite{Nesting} 
$\epsilon_{\mbox{eff}}$  is the effective next-nearest neighbor 
hopping in the 
diagonalized Hamiltonian, while $t_b'$ takes over the role of an effective 
nearest neighbor hopping, i. e.  it does not affect the perfect nesting  
for SDW$_+$ and  SDW$_-$. This is in contrast to the interband SDW$_0$ 
ordering for which finite $t_b'$  introduces an imperfect nesting, and 
$t_b$ keeps the nesting perfect. For temperatures $T>\epsilon_{\mbox{eff}}$  
the fine plateau on Fig.\ref{omega1} is washed away and rounded to a single 
maximum at $p= \pi/d$ and $k_1 \approx 2V(1+t_b^2/V^2)/v_F$ for $V\gg t_b$. 

The magnetic pattern of the phase SDW$_-$ is represented on Fig.\ref{abab}(c) 
for $\eta\approx 0.3$.  As $V$ increases the parameter 
$\eta$  approaches the value $1/2$ from below, 
i. e. the  SDW$_-$ modulation 
in the transverse direction is closer and closer to the antiferromagnetic form 
from Fig.\ref{abab}(b). Furthermore, in the limit $t_b/V \rightarrow 0$
the nonlogarithmic term in eq.(\ref{res})
reduces to
$ 
 \chi_{\mbox{\small res}}(\pm k_1,\pi/d) \approx 
\frac{t_b^2}{V^2} \biggl[ \Psi\biggl(\frac{1}{2}\biggr) -  
\Re \Psi \biggl( \frac{1}{2} + 
\frac{iV}{2\pi T}\biggr)\biggr],
\label{rest}
$
i. e.  the critical temperature approaches its asymptotic value for the 
perfect 
nesting, $T_c^0$. Note that in this limit $\chi_{\mbox{\small res}}(0,0)$ 
tends to $\Psi(\frac{1}{2}) -  \Re \Psi ( \frac{1}{2} + \frac{iV}{2\pi T})$, 
as is seen from eq.(\ref{resv}), i. e. on Fig. \ref{omega1}a the peaks at 
${\bf q}_{\pm}$ dominate over that at ${\bf q}_0$.

\begin{figure}[htbp]
  \begin{center}
    \setlength{\unitlength}{1cm}
    \begin{picture}(9,11)
       \put(-4.2,-6){\includegraphics[width=18cm]{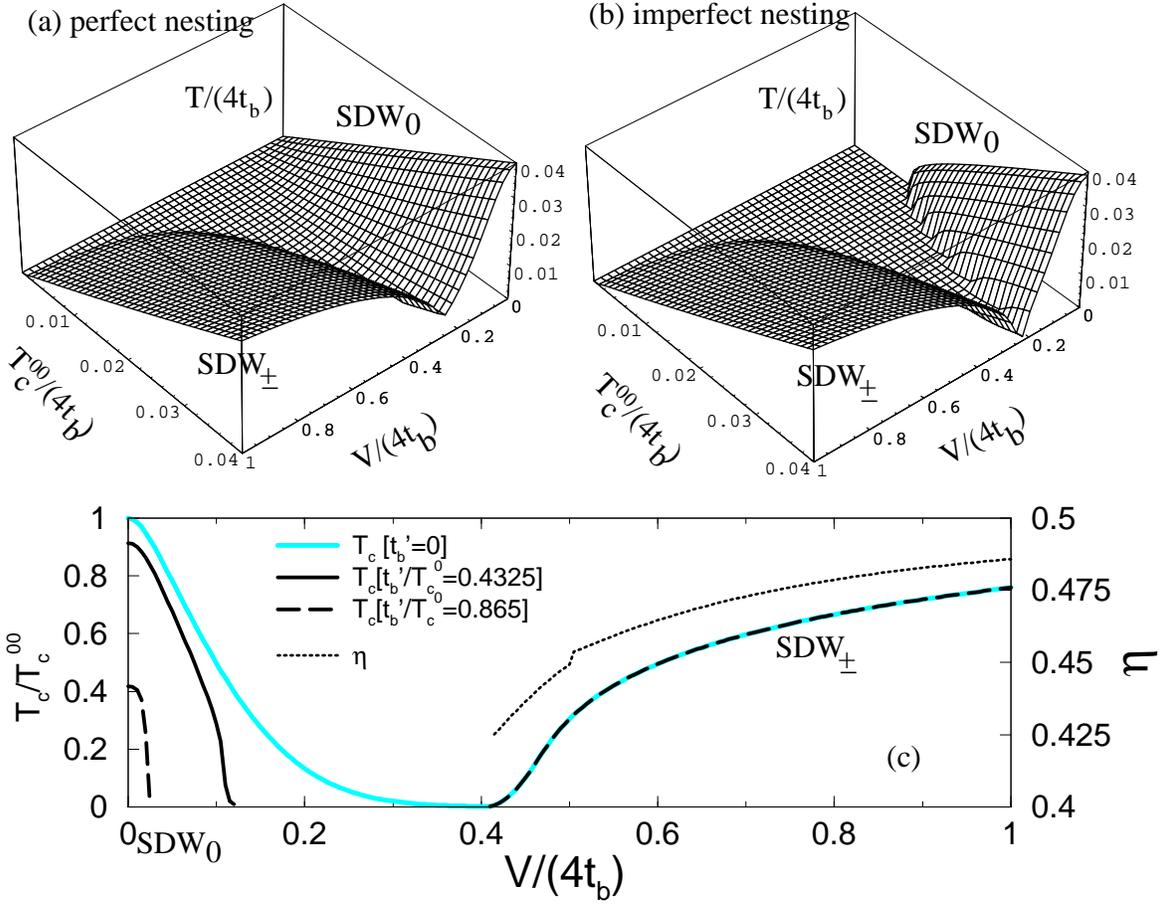}}
    \end{picture}
    \caption{The phase diagram. Small discontinuity in the $\eta$ {\em vs} 
    $V/4t_b$ dependence is caused by the jump from $k^+_2$ to 
    $(k^+_1+k^+_2)/2$ in the Stoner criterion (\ref{stoner}).}
    \label{phd}
  \end{center}
\end{figure}
The phase diagram that results from the above considerations is shown in 
Fig.\ref{phd} in which the Coulomb coupling strength is parametrized by the 
perfect nesting critical temperature ${T_c^{0}}$. Fig.\ref{phd}a
represents the case with $t_b'=0$. Then the low-$V$ (SDW$_0$) and high-$V$ 
(SDW$_{\pm}$) phases are clearly separated by a "valley" in the range of 
intermediate values of $V$. The case with the imperfect nesting is 
illustrated on Fig.\ref{phd}(b) for $t_b'=0.0375t_b$. 

The representatives of $T_c(V)$ dependences for the perfect nesting
and highly imperfect nesting cases, as well as the $V$-dependence of the
parameter $\eta$, are shown on Fig.\ref{phd}(c). 
The Coulomb interaction is chosen to give $T_c^0=0.0108\times 4t_b$
(i. e.  $T_c^0=13K$ for $t_b=300K$). As was already pointed out, 
the critical temperature for SDW$_\pm$ phases, and the parameter  $\eta$ as 
well [right side of Fig.\ref{phd}(c)] are not affected by $t_b'$. On contrary,
the critical temperature for phase SDW$_0$ is very sensitive on the parameter 
$t_b'$, and is completely suppressed for $t_b'\approx T_c^{0}$.

An attempt to make the correspondence of our phase diagram with the experimental
one for $(TMTSF)_2ClO_4$ cannot be pursued without ambiguities. The main problem
is a reliable choice of the value for the parameter  $t_b'$ (beside the 
more or less well established values for other  parameters, $t_b =300K$ and 
$T_c^0=13K$. The value $t_b'=11.25K$ is suggested from the experiments in 
quenched system \cite{Qualls00} under the assumption that the sample is 
entirely free from  anion ordering, i. e. that $V=0$. 
Since one cannot exclude that even for the fastest quenchings  some 
residual anion ordering with a finite, presumably small, value of anion 
potential $V=V_{\mbox{res}}>0$ remains, the above estimate gives an upper 
limit for $t_b'$ in (TMTSF)$_2$ClO$_4$.

Choosing this value of $t_b'$, assuming that $V=0$ in the quenched samples, 
and expecting that $V$ increases monotonously with the cooling rate (e. g. in 
measurements from Refs. \cite{Qualls00,Matsunaga}), we come to the dashed line 
on Fig. \ref{phd}(c) as a fitting curve for (TMTSF)$_2$ClO$_4$. This enables 
the estimation of the lower and upper limit for $V$ in the relaxed salt. The 
latter follow from the width of the region where SDW phases do not exist,
$
0.1 < V(\mbox{relaxed }ClO_4)/t_b < 1.6
$.

In conclusion, we have shown that the anion ordering in Q1D metals 
have some nontrivial consequences even in zero magnetic field.
Fundamentally new concepts in this context are the need for a matrix SDW 
susceptibility in solving the interacting problem, and the
subsequent introduction of the {\em effective bare susceptibilities} that
diagonalize this matrix. Namely,  all previous approaches to the interacting 
system with anion ordering used inadequately the Stoner criterion with
simple SDW bare magnetic susceptibility \cite{Maki86,Gorkov}, or, 
equivalently, 
assumed that the order parameter is scalar in the bond--antibond 
space \cite{Kishigi_etc,Miyazaki98}. Here we showed that the
non-diagonal kinetic energy in bond--antibond representation
is the one that dictates which effective bare susceptibility 
defines the Stoner 
criterion (\ref{stoner}). The novel proposition of the present work is the 
existence of the hybride SDW$_{\pm}$ phase at intermediate values of anion 
potential. Moreover, we have estimated precisely the lower and the
upper limit of $V$ for $(TMTSF)_2ClO_4$ salt.

The next important question is how the magnetic field influences the 
susceptibilities $\tilde{\chi}_{1,2}$ given by eq. (\ref{tilde}). Since the 
present matrix RPA is valid for finite magnetic fields as well, it remains to 
take properly into account the qualitatively more complex content of the
kinematic part of the Hamiltonian (\ref{hint},\ref{hkin}). We leave this part 
of work for future, expecting that the basic physics of the SDW phase in high 
magnetic
fields is indeed that of two  instabilities, as proposed already
by McKernan et al \cite{McKernan95}). More specificaly, our analysis
suggests that two relevant instabilities could be those appearing  at 
wavevectors 
 $2k_F{\bf \hat{a}}+{\bf q}_+$ and $2k_F{\bf \hat{a}}+{\bf q}_-$ 
 on Fig. \ref{omega1}. 

{\bf Acknowledgments:}
 A. B. acknowledges the support of Universit\'{e}
Paris VII during his stay at LPTHE. 
LPTHE is supported by CNRS as Unit\'e Mixte de Recherche, UMR7589.


\end{document}